\documentclass[12pt,a4paper,twocolumn]{article}

\usepackage{subfigure}
\usepackage{widetext}
\usepackage{ifthen} 
\newboolean{pdflatex}
\setboolean{pdflatex}{true} 

\newboolean{articletitles}
\setboolean{articletitles}{true} 

\newboolean{uprightparticles}
\setboolean{uprightparticles}{false} 

\newboolean{inbibliography}
\setboolean{inbibliography}{false} 

\usepackage{microtype}
\usepackage{lineno}  
\usepackage{xspace} 

\usepackage{graphicx}  
\usepackage{color}
\usepackage{colortbl}
\graphicspath{{./figs/}} 

\usepackage{amsmath} 
\usepackage{amssymb}
\usepackage{amsfonts}
\usepackage{upgreek} 

\newcommand*\patchAmsMathEnvironmentForLineno[1]{%
\expandafter\let\csname old#1\expandafter\endcsname\csname #1\endcsname
\expandafter\let\csname oldend#1\expandafter\endcsname\csname
end#1\endcsname
 \renewenvironment{#1}%
   {\linenomath\csname old#1\endcsname}%
   {\csname oldend#1\endcsname\endlinenomath}%
}
\newcommand*\patchBothAmsMathEnvironmentsForLineno[1]{%
  \patchAmsMathEnvironmentForLineno{#1}%
  \patchAmsMathEnvironmentForLineno{#1*}%
}
\AtBeginDocument{%
\patchBothAmsMathEnvironmentsForLineno{equation}%
\patchBothAmsMathEnvironmentsForLineno{align}%
\patchBothAmsMathEnvironmentsForLineno{flalign}%
\patchBothAmsMathEnvironmentsForLineno{alignat}%
\patchBothAmsMathEnvironmentsForLineno{gather}%
\patchBothAmsMathEnvironmentsForLineno{multline}%
}

\usepackage{hyperref}    
\usepackage[all]{hypcap} 

\def\lhcb {\mbox{LHCb}\xspace}

\ifthenelse{\boolean{uprightparticles}}%
{

 \def\Pmu         {\ensuremath{\upmu}\xspace}

 \def\Ppsi        {\ensuremath{\uppsi}\xspace}

 \def\PDelta      {\ensuremath{\Delta}\xspace}                 
 \def\PXi      {\ensuremath{\Xi}\xspace}                 
 \def\PLambda      {\ensuremath{\Lambda}\xspace}                 
 \def\PSigma      {\ensuremath{\Sigma}\xspace}                 
 \def\POmega      {\ensuremath{\Omega}\xspace}                 
 \def\PUpsilon      {\ensuremath{\Upsilon}\xspace}                 
 
 \def\PB      {\ensuremath{\mathrm{B}}\xspace}                 
                  
 \def\PD      {\ensuremath{\mathrm{D}}\xspace}

 \def\PJ      {\ensuremath{\mathrm{J}}\xspace}                 
 \def\PK      {\ensuremath{\mathrm{K}}\xspace}

 \def\Pb      {\ensuremath{\mathrm{b}}\xspace}                 
 \def\Pc      {\ensuremath{\mathrm{c}}\xspace}

 \def\Pi      {\ensuremath{\mathrm{i}}\xspace}

 \def\Ps      {\ensuremath{\mathrm{s}}\xspace}

}
{

 \def\Pmu         {\ensuremath{\mu}\xspace}

 \def\Ppsi        {\ensuremath{\psi}\xspace}                 
                  
 \mathchardef\PDelta="7101
 \mathchardef\PXi="7104
 \mathchardef\PLambda="7103
 \mathchardef\PSigma="7106
 \mathchardef\POmega="710A
 \mathchardef\PUpsilon="7107
                  
 \def\PB      {\ensuremath{B}\xspace}                 
                  
 \def\PD      {\ensuremath{D}\xspace}

 \def\PJ      {\ensuremath{J}\xspace}                 
 \def\PK      {\ensuremath{K}\xspace}

 \def\Pb      {\ensuremath{b}\xspace}                 
 \def\Pc      {\ensuremath{c}\xspace}

 \def\Pi      {\ensuremath{i}\xspace}

 \def\Ps      {\ensuremath{s}\xspace}

}

\def\mup        {\ensuremath{\Pmu^+}\xspace}
\def\mun        {\ensuremath{\Pmu^-}\xspace} 
\def\mumu       {\ensuremath{\Pmu^+\Pmu^-}\xspace}

\def\squark    {\ensuremath{\Ps}\xspace}

\def\cquark    {\ensuremath{\Pc}\xspace}

\def\bquark    {\ensuremath{\Pb}\xspace}

\def\kaon  {\ensuremath{\PK}\xspace}
  \def\Kbar  {\kern 0.2em\overline{\kern -0.2em \PK}{}\xspace}

\def\Kstarz  {\ensuremath{\kaon^{*0}}\xspace}
\def\Kstarzb {\ensuremath{\Kbar^{*0}}\xspace}

  \def\Dbar    {\kern 0.2em\overline{\kern -0.2em \PD}{}\xspace}

\def\B       {\ensuremath{\PB}\xspace}
\def\Bbar    {\ensuremath{\kern 0.18em\overline{\kern -0.18em \PB}{}}\xspace}

\def\Bz      {\ensuremath{\B^0}\xspace}
\def\Bzb     {\ensuremath{\Bbar^0}\xspace}

\def\Bd      {\ensuremath{\B^0}\xspace}
\def\Bs      {\ensuremath{\B^0_\squark}\xspace}

\def\jpsi     {\ensuremath{{\PJ\mskip -3mu/\mskip -2mu\Ppsi\mskip 2mu}}\xspace}
\def\psitwos  {\ensuremath{\Ppsi{(2S)}}\xspace}

  \def\Y#1S{\ensuremath{\PUpsilon{(#1S)}}\xspace}

\def\L {\ensuremath{\PLambda}\xspace}
\def\Lbar {\ensuremath{\kern 0.1em\overline{\kern -0.1em\PLambda}}\xspace}

\def\Lb      {\ensuremath{\L^0_\bquark}\xspace}

\newcommand{\decay}[2]{\ensuremath{#1\!\to #2}\xspace}         

\def\to                 {\ensuremath{\rightarrow}\xspace}

\def\BdToKstmm    {\decay{\Bd}{\Kstarz\mup\mun}}

\def\BdToJPsiKst  {\decay{\Bd}{\jpsi\Kstarz}}

\def\AT#1     {\ensuremath{A_{\mathrm{T}}^{#1}}\xspace}           

\def\C#1      {\ensuremath{\mathcal{C}_{#1}}\xspace}                       
\def\Cp#1     {\ensuremath{\mathcal{C}_{#1}^{'}}\xspace}                    
\def\Ceff#1   {\ensuremath{\mathcal{C}_{#1}^{\mathrm{(eff)}}}\xspace}        
\def\Cpeff#1  {\ensuremath{\mathcal{C}_{#1}^{'\mathrm{(eff)}}}\xspace}       
\def\Ope#1    {\ensuremath{\mathcal{O}_{#1}}\xspace}                       
\def\Opep#1   {\ensuremath{\mathcal{O}_{#1}^{'}}\xspace}                    

\newcommand{\tev}{\ifthenelse{\boolean{inbibliography}}{\ensuremath{~T\kern -0.05em eV}\xspace}{\ensuremath{\mathrm{\,Te\kern -0.1em V}}\xspace}}
\newcommand{\gev}{\ensuremath{\mathrm{\,Ge\kern -0.1em V}}\xspace}
\newcommand{\mev}{\ensuremath{\mathrm{\,Me\kern -0.1em V}}\xspace}
\newcommand{\kev}{\ensuremath{\mathrm{\,ke\kern -0.1em V}}\xspace}
\newcommand{\ev}{\ensuremath{\mathrm{\,e\kern -0.1em V}}\xspace}
\newcommand{\gevc}{\ensuremath{{\mathrm{\,Ge\kern -0.1em V\!/}c}}\xspace}
\newcommand{\mevc}{\ensuremath{{\mathrm{\,Me\kern -0.1em V\!/}c}}\xspace}
\newcommand{\gevcc}{\ensuremath{{\mathrm{\,Ge\kern -0.1em V\!/}c^2}}\xspace}
\newcommand{\gevgevcccc}{\ensuremath{{\mathrm{\,Ge\kern -0.1em V^2\!/}c^4}}\xspace}
\newcommand{\mevcc}{\ensuremath{{\mathrm{\,Me\kern -0.1em V\!/}c^2}}\xspace}

\def\mum  {\ensuremath{\,\upmu\rm m}\xspace}

\def\invfb   {\ensuremath{\mbox{\,fb}^{-1}}\xspace}

\def\gsim{{~\raise.15em\hbox{$>$}\kern-.85em
          \lower.35em\hbox{$\sim$}~}\xspace}
\def\lsim{{~\raise.15em\hbox{$<$}\kern-.85em
          \lower.35em\hbox{$\sim$}~}\xspace}

\def\pt         {\mbox{$p_{\rm T}$}\xspace}

\def\evtgen     {\mbox{\textsc{EvtGen}}\xspace}

\def\geant      {\mbox{\textsc{Geant4}}\xspace}

\def\photos     {\mbox{\textsc{Photos}}\xspace}

\def\pythia     {\mbox{\textsc{Pythia}}\xspace}

\def\tell1  {TELL1\xspace}
\def\ukl1   {UKL1\xspace}

\usepackage{cite} 
\usepackage{mciteplus}

\begin{document}

\renewcommand{\thefootnote}{\fnsymbol{footnote}}
\setcounter{footnote}{1}

\onecolumn

\begin{titlepage}
\pagenumbering{roman}

\vspace*{-1.5cm}
\centerline{\large EUROPEAN ORGANIZATION FOR NUCLEAR RESEARCH (CERN)}
\vspace*{1.5cm}
\hspace*{-0.5cm}
\begin{tabular*}{\linewidth}{lc@{\extracolsep{\fill}}r}
\ifthenelse{\boolean{pdflatex}}
{\vspace*{-2.7cm}\mbox{\!\!\!\includegraphics[width=.14\textwidth]{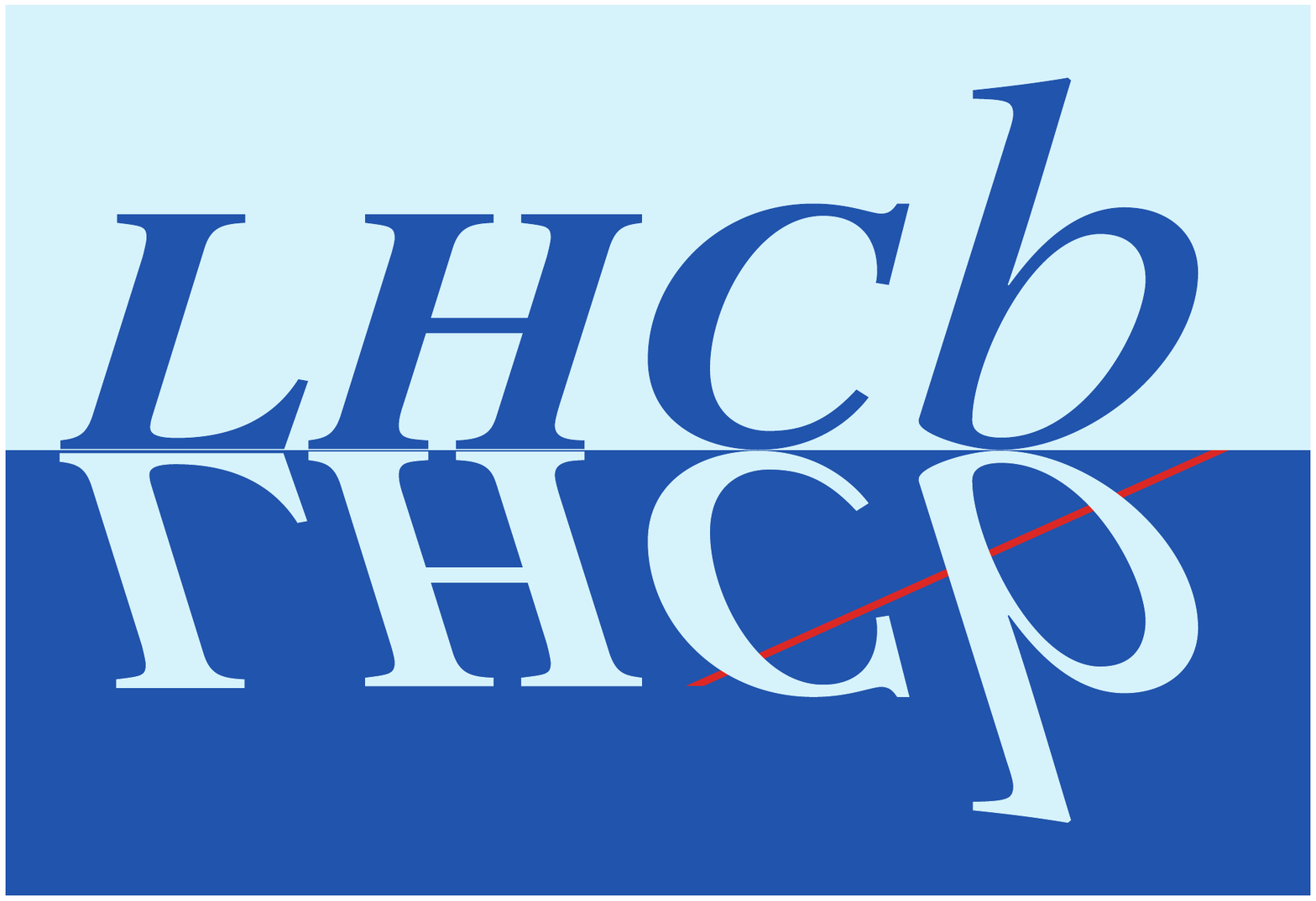}} & &}%
{\vspace*{-1.2cm}\mbox{\!\!\!\includegraphics[width=.12\textwidth]{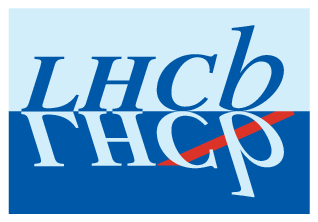}} & &}%
\\
 & & CERN-PH-EP-2013-146 \\  
 & & LHCb-PAPER-2013-037 \\  
 & & 7 August 2013 \\ 
 & & \\
\end{tabular*}

\vspace*{2.5cm}

{\bf\boldmath\huge
\begin{center}
  Measurement of form-factor independent observables in the decay $B^{0} \to K^{*0} \mu^+ \mu^-$
\end{center}
}

\vspace*{2.0cm}

\begin{center}
The LHCb collaboration\footnote{Authors are listed on the following pages.}
\end{center}

\vspace{\fill}

\begin{abstract}
  \noindent
\noindent We present a measurement of form-factor independent angular observables in the decay \mbox{$B^0\to K^{*}(892)^{0}\mu^+ \mu^-$}.  
The analysis is based on a data sample corresponding to an
integrated luminosity of 1.0\invfb, collected by the LHCb experiment
in $pp$ collisions at a center-of-mass energy of 7\tev. Four observables are measured
in six bins of the dimuon invariant mass squared, $q^2$, in the range $0.1<q^2<19.0$\gevgevcccc. Agreement
with Standard Model predictions is found for 23 of the 24 measurements. 
A local discrepancy, corresponding to $3.7$ Gaussian standard deviations, is observed in one $q^2$ bin for 
one of the observables. Considering the 24 measurements as independent, 
the probability to observe such a discrepancy, or larger, in one is $0.5\%$. 

\end{abstract}
\vspace*{2.0cm}

\begin{center}
  Submitted to Phys.~Rev.~Lett. 
\end{center}

\vspace{\fill}

{\footnotesize 
\centerline{\copyright~CERN on behalf of the \lhcb collaboration, license \href{http://creativecommons.org/licenses/by/3.0/}{CC-BY-3.0}.}}
\vspace*{2mm}

\end{titlepage}


\newpage
\setcounter{page}{2}
\mbox{~}
\newpage

\centerline{\large\bf LHCb collaboration}
\begin{flushleft}
\small
R.~Aaij$^{40}$, 
B.~Adeva$^{36}$, 
M.~Adinolfi$^{45}$, 
C.~Adrover$^{6}$, 
A.~Affolder$^{51}$, 
Z.~Ajaltouni$^{5}$, 
J.~Albrecht$^{9}$, 
F.~Alessio$^{37}$, 
M.~Alexander$^{50}$, 
S.~Ali$^{40}$
% ...

{\footnotesize \it
% ...
$ ^{5}$Clermont Universit\'{e}, Universit\'{e} Blaise Pascal, CNRS/IN2P3, LPC, Clermont-Ferrand, France\\
$ ^{6}$CPPM, Aix-Marseille Universit\'{e}, CNRS/IN2P3, Marseille, France\\
% ...
$ ^{9}$Fakult\"{a}t Physik, Technische Universit\"{a}t Dortmund, Dortmund, Germany\\
% ...
$ ^{36}$Universidad de Santiago de Compostela, Santiago de Compostela, Spain\\
$ ^{37}$European Organization for Nuclear Research (CERN), Geneva, Switzerland\\
% ...
$ ^{40}$Nikhef National Institute for Subatomic Physics, Amsterdam, The Netherlands\\
% ...
$ ^{45}$H.H. Wills Physics Laboratory, University of Bristol, Bristol, United Kingdom\\
% ...
$ ^{50}$School of Physics and Astronomy, University of Glasgow, Glasgow, United Kingdom\\
$ ^{51}$Oliver Lodge Laboratory, University of Liverpool, Liverpool, United Kingdom\\
% ...
}
\end{flushleft}

\cleardoublepage

\twocolumn

\renewcommand{\thefootnote}{\arabic{footnote}}
\setcounter{footnote}{0}



\pagestyle{plain} 
\setcounter{page}{1}
\pagenumbering{arabic}


The rare decay $B^0\to K^{*0} \mu^+ \mu^-$,
where \Kstarz indicates the $K^{*}(892)^{0}\to K^+ \pi^-$ decay, 
is a flavor-changing neutral current process that proceeds via 
loop and box amplitudes in the Standard Model (SM). In extensions of the SM, contributions from new particles 
can enter in competing amplitudes and modify the angular distributions of the decay products. 
This decay has been widely studied from both theoretical~\cite{Altmannshofer:2008dz,Becirevic:2011bp,Matias:2012xw} and 
experimental~\cite{Aubert:2008ju,Wei:2009zv,Aaltonen:2011ja,Aaij:2013iag} perspectives. 
Its angular distribution is described by three angles ($\theta_{\ell}$, $\theta_K$ and $\phi$) and the 
dimuon invariant mass squared, $q^2$; $\theta_\ell$ is the angle between the flight direction of the \mup (\mun) and the \Bz (\Bzb) meson in the dimuon rest frame; 
$\theta_K$ is the angle between the flight direction of the charged kaon and the \Bz (\Bzb) meson in the \Kstarz (\Kstarzb) rest frame; and $\phi$ is the angle between the decay planes of the \Kstarz (\Kstarzb) and the dimuon system in the \Bz (\Bzb) meson rest frame. 
A formal definition of the angles can be found in Ref.~\cite{Aaij:2013iag}. 
Using the definitions of Ref.~\cite{Altmannshofer:2008dz} and summing over \Bz and \Bzb mesons, 
the differential angular distribution can be written as 

\begin{widetext}
\begin{equation}
\begin{split}
\frac{1}{\mathrm{d}\Gamma/dq^2} \frac{\mathrm{d}^{4}\Gamma}{\mathrm{d}\cos\theta_{\ell}\,\mathrm{d}\cos\theta_{K}\,\mathrm{d}\phi\,\mathrm{d}q^2}
= & \frac{9}{32\pi} \left[ \frac{3}{4}(1-F_{\rm L}) \sin^2 \theta_K + F_{\rm L}
  \cos^2 \theta_K+ \frac{1}{4}(1- F_{\rm L}) \sin^2 \theta_K \cos 2\theta_{\ell} \right. \\ &-
  \left. F_{\rm L} \cos^2 \theta_K \cos 2\theta_{\ell}  + S_{3} \sin^2 \theta_K \sin^{2} \theta_{\ell}
  \cos 2\phi \right. \\ &
 \left. +~ S_{4} \sin 2\theta_K \sin 2\theta_{\ell} \cos\phi \right. 
\left. +~ S_{5} \sin 2\theta_K \sin\theta_{\ell}\cos\phi \right.  \\  & 
\left. +~ S_{6} \sin^2 \theta_K \cos\theta_{\ell} +  S_{7} \sin 2\theta_K \sin\theta_{\ell} \sin\phi  \right. \\  &
\left. +~ S_{8} \sin
  2\theta_K \sin 2\theta_{\ell}\sin\phi + S_{9}
  \sin^2 \theta_K \sin^{2}\theta_{\ell} \sin 2\phi \frac{}{}~\right],
\end{split}\label{eq:masterformula}
\end{equation}  
\end{widetext}

\noindent where the $q^2$ dependent observables $F_{\rm L}$ and $S_i$ are bilinear combinations of the \Kstarz
decay amplitudes. These in turn are functions of the Wilson
coefficients, which contain information about short distance effects and are sensitive to physics beyond the SM, and form-factors, which 
depend on long distance effects. 
Combinations of  $F_{\rm L}$ and $S_i$ with reduced form-factor uncertainties
have been proposed independently by several authors~\cite{Kruger:2005ep,Egede:2008uy,Becirevic:2011bp,Matias:2012xw,Bobeth:2011gi}. 
In particular, in the large recoil limit (low-$q^2$) the
observables denoted as $P_{4}^{\prime}$, $P_{5}^{\prime}$, $P_{6}^{\prime}$ and $P_{8}^{\prime}$~\cite{Descotes-Genon:2013vna} 
are largely free from
form-factor uncertainties. These observables are defined as 
\begin{equation}
P_{i=4,5,6,8}^{\prime} = \frac{S_{j=4,5,7,8}}{\sqrt{F_{\rm L} (1-F_{\rm L})}}.
\end{equation}

This Letter presents the measurement of the observables $S_j$ and the
respective observables $P_i^{\prime}$. 
This is the first measurement of these quantities by any experiment. Moreover, these observables 
provide complementary information about physics beyond the SM with respect to the angular observables 
previously measured in this decay~\cite{Aubert:2008ju,Wei:2009zv,Aaltonen:2011ja,Aaij:2013iag}. 
The data sample analyzed corresponds to an integrated luminosity
of 1.0\invfb of $pp$ collisions at a center-of-mass energy of 7 TeV collected by the LHCb experiment in 2011. 
Charged conjugation is implied throughout this Letter, unless otherwise stated.

The \lhcb detector~\cite{Alves:2008zz} is a single-arm forward
spectrometer covering the \mbox{pseudorapidity} range $2<\eta <5$,
designed for the study of particles containing \bquark or \cquark
quarks. The detector includes a high-precision tracking system
consisting of a silicon-strip vertex detector surrounding the $pp$
interaction region, a large-area silicon-strip detector located
upstream of a dipole magnet with a bending power of approximately 
$4{\rm\,Tm}$, and three stations of silicon-strip detectors and straw
drift tubes placed downstream of the magnet.
The combined tracking system provides a momentum measurement with
relative uncertainty that varies from 0.4\% at 5\gevc to 0.6\% at 100\gevc,
and impact parameter resolution of 20\mum for
tracks with high transverse momentum ($\pt$). Charged hadrons are identified
using two ring-imaging Cherenkov detectors~\cite{Adinolfi:1495721}. 
Muons are identified by a
system composed of alternating layers of iron and multiwire
proportional chambers~\cite{AlvesJr:1492807}.

The trigger~\cite{Aaij:2012me} consists of a
hardware stage, based on information from the calorimeter and muon
systems, followed by a software stage, which applies a full event
reconstruction.
Candidate events for this analysis are required to pass a hardware trigger, 
which selects muons with $\pt>1.48\gevc$. 
In the software trigger, at least one of the final 
state particles is required to have both $\pt>1.0\gevc$ and 
impact parameter larger than $100\mum$ with respect to all of the primary $pp$ 
interaction vertices in the event. 
Finally, the tracks of two or more of the final state particles 
are required to form a vertex that is significantly displaced from the primary vertex.

Simulated events are used in several stages of the analysis, $pp$ collisions are generated using
\pythia~6.4~\cite{Sjostrand:2006za} with a specific \lhcb
configuration~\cite{LHCb-PROC-2010-056}.  Decays of hadronic particles
are described by \evtgen~\cite{Lange:2001uf}, in which final state
radiation is generated using \photos~\cite{Golonka:2005pn}. The
interaction of the generated particles with the detector and its
response are implemented using the \geant
toolkit~\cite{Allison:2006ve, *Agostinelli:2002hh} as described in
Ref.~\cite{LHCb-PROC-2011-006}. 

This analysis uses the same selection and acceptance correction
technique as described in Ref.~\cite{Aaij:2013iag}. 

Signal candidates are required to
pass a loose preselection: the \Bz
vertex is required to be well separated from the primary $pp$
interaction point; the impact
parameter with respect to the primary $pp$ interaction point is
required to be small for the
\Bz candidate and large for the final state
particles; and the angle between the \Bz momentum and the vector
from the primary vertex to the \Bz decay vertex is required to be small. 
Finally, the reconstructed invariant mass of the \Kstarz candidate is required 
to be in the range \mbox{$792<m_{K\pi}<992$\mevcc}. 
To further reject combinatorial background events, a boosted decision tree (BDT) ~\cite{Breiman}
using the AdaBoost algorithm~\cite{AdaBoost}  is applied.
The BDT combines kinematic and geometrical properties of the
event. 

Several sources of peaking background have been considered. 
The decays \BdToJPsiKst and \decay{\Bz}{\psitwos\Kstarz}, where the
charmonium resonances decay into a muon pair, are rejected by vetoing events for which the
dimuon system has an invariant mass ($m_{\mu \mu}$) in the range
$2946-3176$\mevcc or $3586-3766$\mevcc. 
Both vetoes are extended downwards by 150\mevcc for
\Bz candidates with invariant mass ($m_{K\pi\mu\mu}$) in the range 
$5150-5230$\mevcc to account for the
radiative tails of the charmonium resonances. They are also extended
upwards by 25\mevcc for candidates with $5370<m_{K\pi\mu\mu}<5470$\mevcc,
to account for non-Gaussian reconstruction effects. 
Backgrounds from \BdToJPsiKst decays with the kaon
or pion from the \Kstarz decay and one of the muons from the $J/\psi$ meson being misidentified and 
swapped with each other, are rejected by assigning the muon mass hypothesis to the $K^+$ or $\pi^-$ and vetoing candidates for which 
the resulting invariant mass is in the range $3036<m_{\mu\mu}<3156$\mevcc. 
Background from $\Bs \to \phi (\to K^+ K^-) \mu^+\mu^-$ decays is removed by 
assigning the kaon mass hypothesis to the pion candidate and rejecting events for which the 
resulting invariant mass $K^+ K^-$ is consistent with the $\phi$ mass. 
A similar veto is applied to remove $\Lb \to\L (1520) (\to p K^-) \mu^+ \mu^-$ events. 
After these vetoes, the remaining peaking background is estimated to be
negligibly small. It has been verified with the simulation that these vetos do not bias the angular observables. 
In total, 883 signal candidates are observed in the 
range $0.1<q^2<19.0$\gevgevcccc , with a signal over background ratio of about 5. 

Detector acceptance effects are accounted for by weighting the
candidates with the inverse of their efficiency. 
The efficiency is determined as a function of the three angles and
$q^2$ by using a large sample of simulated events and assuming factorization in the
three angles. Possible non-factorizable acceptance effects are evaluated and included in the systematic uncertainties. 
Several control channels,
in particular the decay \BdToJPsiKst, which has the same
final state as the signal, are used to verify the agreement between data and
simulation. 

Due to the limited number of signal candidates in this
dataset, we do not fit the data to the full differential distribution 
of Eq.~\ref{eq:masterformula}. In Ref.~\cite{Aaij:2013iag}, 
the data were ``folded'' at $\phi=0$ ($\phi\to \phi + \pi$ for $\phi<0$) to reduce the number of parameters in the fit, 
while cancelling the terms containing $\sin{\phi}$ and $\cos{\phi}$. 
Here, similar folding techniques are applied to specific regions of the three-dimensional angular space to exploit the (anti)-symmetries 
of the differential decay rate with respect to combinations of angular variables.
This simplifies the differential decay rate without losing experimental sensitivity. 
This technique is discussed in more detail in Ref.~\cite{michelThesis}.  
The following sets of transformations are used to determine the observables of interest 
\begin{eqnarray}
\text{$P_4^{\prime}$, $S_4$:  }
\begin{cases} \phi \to -\phi & \text{~for~} \phi < 0 \\
\phi \to \pi - \phi  & \text{~for~} \theta_{\ell} > \pi/2  \\
\theta_{\ell} \to \pi - \theta_{\ell} & \text{~for~} \theta_{\ell} > \pi/2,  
\end{cases}
\label{eq:foldingP4}
\end{eqnarray}
\begin{eqnarray}
\text{$P_5^{\prime}$, $S_5$:  }
\begin{cases} \phi \to -\phi & \text{~for~} \phi < 0 \\
\theta_{\ell} \to \pi - \theta_{\ell} & \text{~for~} \theta_{\ell} > \pi/2,  
\end{cases}
\label{eq:foldingP5}
\end{eqnarray}
\begin{eqnarray}
\text{$P_6^{\prime}$, $S_7$:  }
\begin{cases} \phi \to \pi -\phi & \text{~for~} \phi > \pi/2 \\
\phi \to -\pi -\phi & \text{~for~} \phi < -\pi/2 \\
\theta_{\ell} \to \pi - \theta_{\ell} & \text{~for~} \theta_{\ell} > \pi/2 ,
\end{cases}
\label{eq:foldingP6}
\end{eqnarray}
\begin{eqnarray}
\text{$P_8^{\prime}$, $S_8$:  }
\begin{cases} 
\phi \to \pi -\phi & \text{~for~} \phi > \pi/2 \\
\phi \to -\pi -\phi & \text{~for~} \phi < -\pi/2 \\
\theta_K \to \pi - \theta_K & \text{~for~} \theta_{\ell} > \pi/2 \\
\theta_{\ell} \to \pi - \theta_{\ell} & \text{~for~} \theta_{\ell} > \pi/2.  
\end{cases}
\label{eq:foldingP8}
\end{eqnarray}
Each transformation preserves the first five terms and the corresponding $S_i$ term in Eq.~\ref{eq:masterformula}, 
and cancels the other angular terms.
Thus, the resulting angular distributions depend only on $F_{\rm L}$, $S_3$ and 
one of the observables $S_{4,5,7,8}$.

Four independent likelihood fits to the \Bd invariant mass
and the transformed angular distributions are performed to extract the observables
$P_{i}^{\prime}$ and $S_{i}$. The signal invariant mass shape is parametrized with the sum of two Crystal Ball
functions~\cite{Skwarnicki:1986xj}, where the parameters are extracted from the fit to 
\BdToJPsiKst decays in data. The background invariant mass shape is
parametrized with an exponential function, while its angular distribution is parametrized with the direct product of 
three second-order polynomials, dependent on $\phi$, $\cos{\theta_K}$ and $\cos{\theta_{\ell}}$. 
The angular observables $F_{\rm L}$ and $S_3$ are allowed to vary in the angular fit and 
are treated as nuisance parameters in this analysis. Their fit 
values agree with Ref.~\cite{Aaij:2013iag}. 

The presence of a $K^+\pi^-$ system in an S-wave
configuration, due to a non-resonant contribution or to feed-down from $K^{+}\pi^-$ scalar resonances, 
results in additional terms in the differential angular distribution.
Denoting the right-hand side of Eq.~\ref{eq:masterformula} by 
$W_{\rm P}$, the differential decay rate takes the form

\begin{equation}                                                                                                                                          
\begin{split}                                                                                                                                             
(1-F_{\rm S}) W_{\text{P}} + \frac{9}{32\pi}  \left(W_{\rm S}+W_{\rm SP} \right),                                                            
\end{split}\label{eq:masterformula:swave}                                                                                                                 
\end{equation} 
where
\begin{equation}
\begin{split}
W_{\rm S} = \frac{2}{3} F_{\rm S} \sin^2 \theta_{\ell} 
\end{split}\label{eq:masterformula:swave2}
\end{equation}
and $W_{\rm SP}$ is given by
\begin{equation}
\begin{split}
&\left.  \frac{4}{3} A_\mathrm{S}\sin^2\theta_{\ell} \cos\theta_K + A_\mathrm{S}^{(4)} \sin\theta_K \sin 2\theta_{\ell} \cos\phi + \right. \\ & 
\left. A_\mathrm{S}^{(5)} \sin\theta_K \sin\theta_{\ell} \cos\phi + A_\mathrm{S}^{(7)} \sin\theta_K \sin\theta_{\ell} \sin\phi \right. \\ &
\left. + A_{\mathrm{S}}^{(8)}\sin\theta_K\sin 2\theta_{\ell} \sin\phi~.~\right.
\end{split}\label{eq:masterformula:swave3}
\end{equation}

\noindent The factor   
$F_{\rm S}$ is the fraction of the S-wave component in the \Kstarz mass window, 
and $W_{\rm SP}$ contains all the interference terms, $A_\mathrm{S}^{(i)}$, 
of the \mbox{S-wave} with the \Kstarz transversity amplitudes as defined in
Ref.~\cite{Matias:2012qz}. 
In Ref.~\cite{Aaij:2013iag}, $F_{\rm S}$ was measured to be less than
$0.07$ at $68\%$ confidence level. 
The maximum value that the quantities
$A_{\mathrm{S}}^{(i)}$ can assume is a function of $F_{\rm S}$ and $F_{\rm L}$~\cite{Descotes-Genon:2013vna}.
The S-wave contribution is neglected in the fit to data, but its effect is evaluated and assigned as a
systematic uncertainty using pseudo-experiments. 
A large number of pseudo-experiments with $F_{\rm S}=0.07$ and with the interference 
terms set to their maximum allowed values are generated. 
All other parameters, including the angular observables, are set to their measured values in data. 
The pseudo-experiments are fitted ignoring S-wave and interference 
contributions. The corresponding bias in the measurement of the angular
observables is assigned as a systematic uncertainty. 

\begin{table*}[!htb]
{\small
\caption{\small Measurement of the observables $P_{4,5,6,8}^{\prime}$  and $S_{4,5,7,8}$ in the six $q^2$ bins of the analysis. 
For the observables $P_i^{\prime}$ the measurement in the $q^2$-bin $1.0<q^2<6.0$ \gevgevcccc, 
which is the theoretically preferred region at large recoil, is also
reported. The first uncertainty is statistical and the second is 
systematic.}
\begin{tabular}{c|cccc}
  $q^2$[\gevgevcccc]   & $P_4^{\prime}$  & $P_5^{\prime}$ &$P_6^{\prime}$ &$P_8^{\prime}$ \\[1ex]
\hline\\
$\phantom{0}0.10\phantom{0}-\phantom{0}2.00$ & $\phantom{0}\;0.00^{+0.26}_{-0.26}\pm 0.03$ & $\phantom{0}\;0.45^{+0.19}_{-0.22}\pm
0.09$  &$-0.24^{+0.19}_{-0.22}\pm 0.05$ & $-0.06^{+0.28}_{-0.28}\pm
0.02$ \\[1ex]
$\phantom{0}2.00\phantom{0}-\phantom{0}4.30$ &  $-0.37^{+0.29}_{-0.26} \pm 0.08$ &
$\phantom{0}\;0.29^{+0.39}_{-0.38}\pm 0.07$& $\phantom{0}\;0.15^{+0.36}_{-0.38}\pm 0.05$ &
$-0.15^{+0.29}_{-0.28}\pm 0.07$ \\[1ex]
$\phantom{0}4.30\phantom{0}-\phantom{0}8.68$ & $-0.59^{+0.15}_{-0.12} \pm 0.05$ &
$-0.19^{+0.16}_{-0.16}\pm 0.03$& $-0.04^{+0.15}_{-0.15}\pm 0.05$&
$\phantom{0}\;0.29^{+0.17}_{-0.19}\pm 0.03$\\[1ex]
$10.09\phantom{0}-12.90$ &  $-0.46^{+0.20}_{-0.17}\pm 0.03$ &
$-0.79^{+0.16}_{-0.19}\pm 0.19$ & $-0.31^{+0.23}_{-0.22}\pm 0.05$ &
$-0.06^{+0.23}_{-0.22}\pm 0.02$\\[1ex]
$14.18\phantom{0}-16.00$ & $\phantom{0}\;0.09^{+0.35}_{-0.27}\pm 0.04$ &
$-0.79^{+0.20}_{-0.13} \pm 0.18$ & $-0.18^{+0.25}_{-0.24}\pm 0.03$
&$-0.20^{+0.30}_{-0.25}\pm 0.03$\\[1ex]
$16.00\phantom{0}-19.00$ & $-0.35^{+0.26}_{-0.22}\pm 0.03$ & $-0.60^{+0.19}_{-0.16}
\pm 0.09$ & $\phantom{0}\;0.31^{+0.38}_{-0.37}\pm 0.10$ & $\phantom{0}\;0.06^{+0.26}_{-0.27}\pm
0.03$ \\[1ex]
$\phantom{0}1.00\phantom{0}-\phantom{0}6.00$ & $-0.29^{+0.18}_{-0.16}\pm 0.03$& $\phantom{0}\;0.21^{+0.20}_{-0.21}\pm 0.03$& 
$-0.18^{+0.21}_{-0.21} \pm 0.03$ & $\phantom{0}\;0.23^{+0.18}_{-0.19}\pm 0.02$\\[1ex]
\hline 
\multicolumn{5}{c}{} \\
 $q^2$[\gevgevcccc] & $S_4$ & $S_5$ & $S_7$ & $S_8$ \\[1ex]
\hline \\
$\phantom{0}0.10\phantom{0}-\phantom{0}2.00$ & $\phantom{0}\;0.00^{+0.12}_{-0.12}\pm 0.03$ & $\phantom{0}\;0.22^{+0.09}_{-0.10}
\pm 0.04$ & $-0.11^{+0.11}_{-0.11} \pm 0.03$ &
$-0.03^{+0.13}_{-0.12}\pm 0.01$ \\[1ex]
$\phantom{0}2.00\phantom{0}-\phantom{0}4.30$ & $-0.14^{+0.13}_{-0.12}\pm 0.03$ & $\phantom{0}\;0.11^{+0.14}_{-0.13}\pm
0.03$ & $\phantom{0}\;0.06^{+0.15}_{-0.15}\pm 0.02$ & $-0.06^{+0.12}_{-0.12} \pm 0.02$ \\[1ex]
$\phantom{0}4.30\phantom{0}-\phantom{0}8.68$ & $-0.29^{+0.06}_{-0.06} \pm 0.02$ & $-0.09^{+0.08}_{-0.08}
\pm 0.01$ & $-0.02^{+0.07}_{-0.08}\pm 0.04$ & $\phantom{0}\;0.15^{+0.08}_{-0.08}
\pm 0.01$ \\[1ex]
$10.09\phantom{0}-12.90$ & $-0.23^{+0.09}_{-0.08} \pm 0.02$ &
$-0.40^{+0.08}_{-0.10} \pm 0.10$ & $-0.16^{+0.11}_{-0.12} \pm 0.03$ &
$-0.03^{+0.10}_{-0.10}\pm 0.01$ \\[1ex]
$14.18\phantom{0}-16.00$ & $\phantom{0}\;0.04^{+0.14}_{-0.08} \pm 0.01$ &
$-0.38^{+0.10}_{-0.09} \pm 0.09$ & $-0.09^{+0.13}_{-0.14} \pm 0.01$ &
$-0.10^{+0.13}_{-0.12}\pm 0.02$ \\[1ex]
$16.00\phantom{0}-19.00$ & $-0.17^{+0.11}_{-0.09} \pm 0.01$ &
$-0.29^{+0.09}_{-0.08}\pm 0.04$ & $\phantom{0}\;0.15^{+0.16}_{-0.15}\pm 0.03$ &
$\phantom{0}\;0.03^{+0.12}_{-0.12}\pm 0.02$ \\[1ex]
\hline
  \end{tabular}\label{tab:results} }
\end{table*}

\begin{figure}[hbt]
\includegraphics[width=0.9\linewidth]{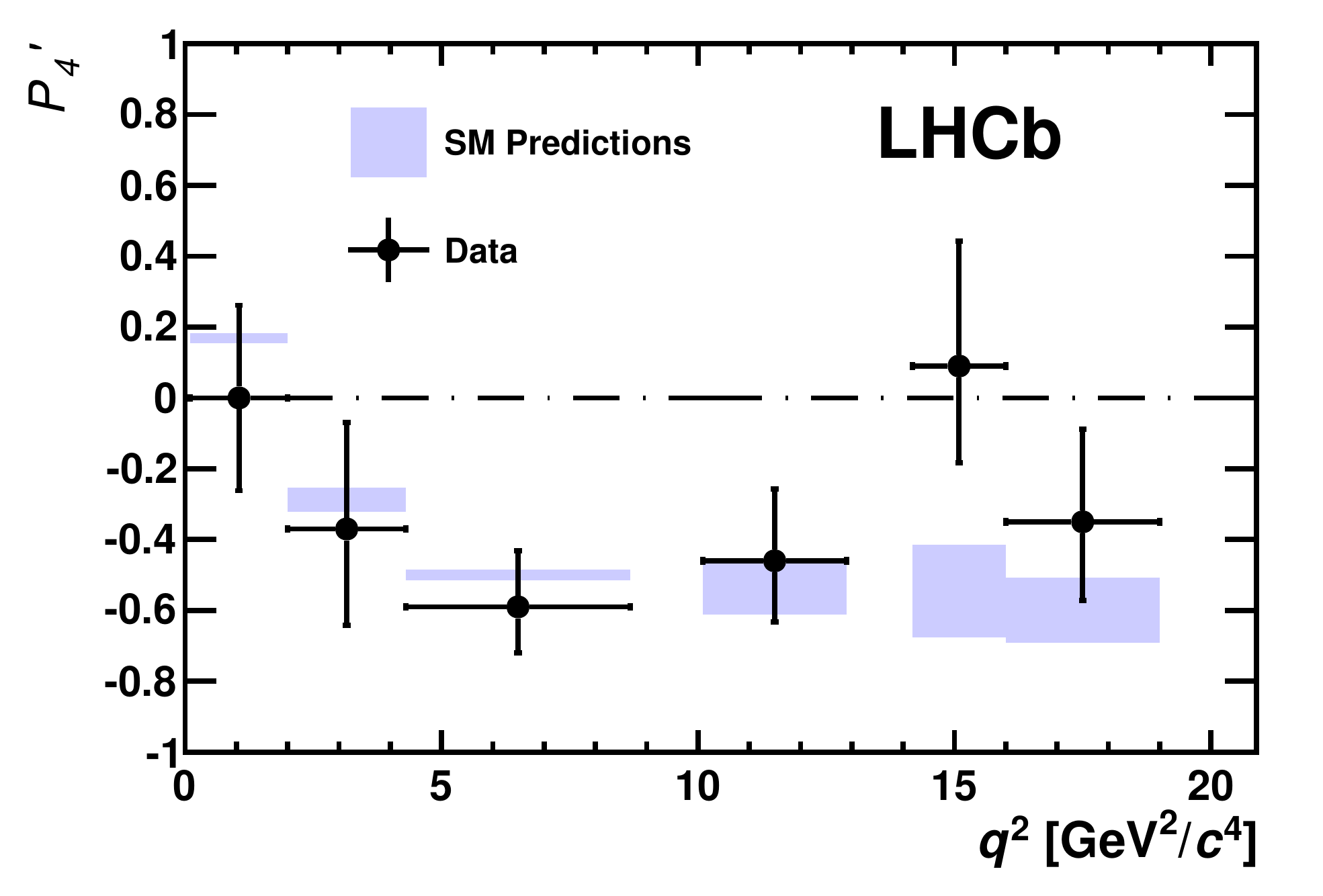}\\
\includegraphics[width=0.9\linewidth]{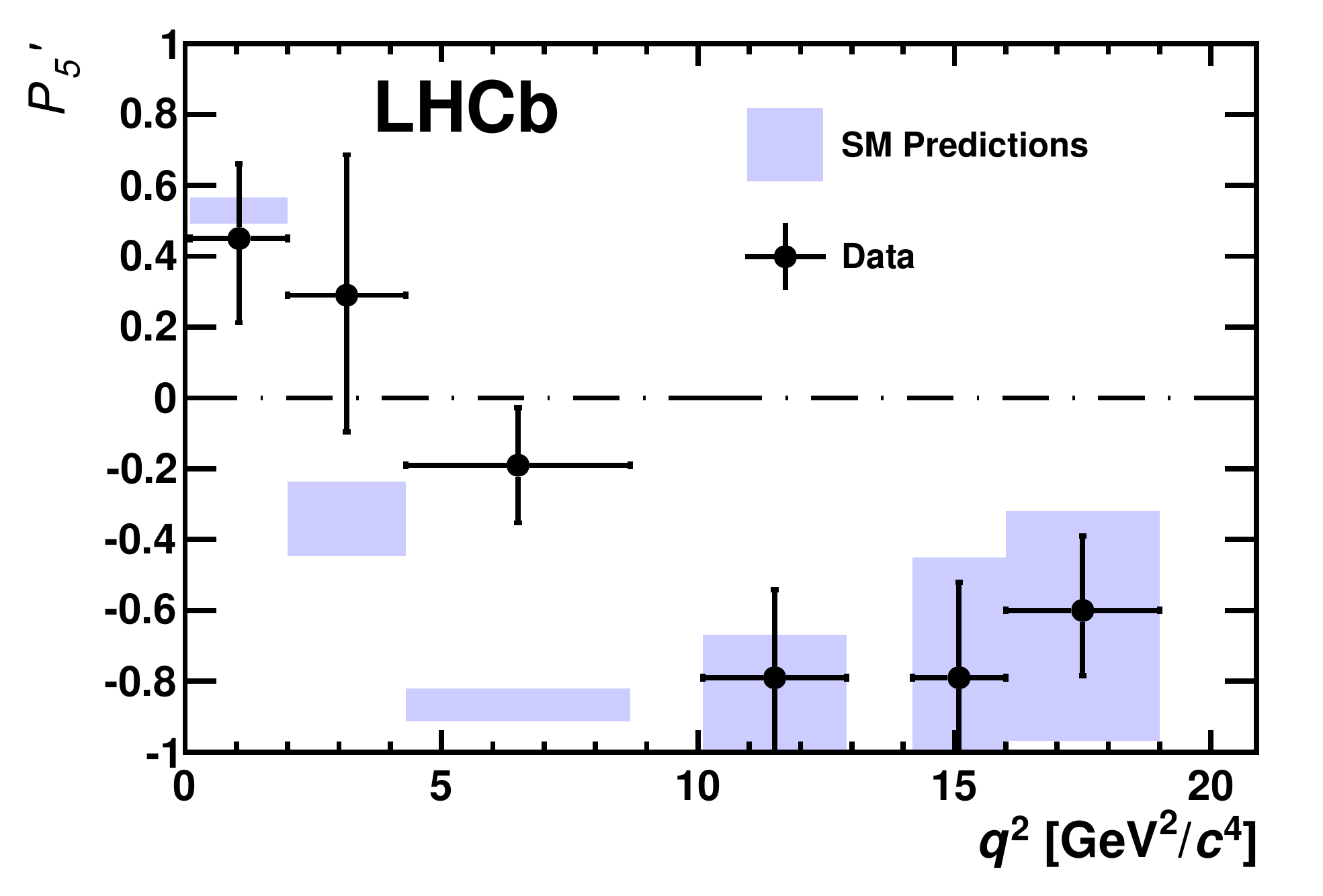}
\caption{\small Measured values of $P_{4}^{\prime}$ and $P_5^{\prime}$
  (black points) compared with SM predictions from Ref.~\cite{Descotes-Genon:2013vna} (blue bands).\label{fig:result:PResult}}
\end{figure}
The results of the angular fits to the data are presented in Table~\ref{tab:results}. 
The statistical uncertainties are determined using the
Feldman-Cousins method~\cite{Feldman:1997qc}. 
The systematic uncertainty takes into account the
limited knowledge of the angular acceptance, uncertainties in the
signal and background invariant mass models, the angular model
for the background, and the impact of a possible S-wave amplitude. 
Effects due to \Bz/\Bzb production asymmetry have been considered and found negligibly small. 
The comparison between the measurements and the
theoretical predictions from Ref.~\cite{Descotes-Genon:2013vna} are shown in Fig.~\ref{fig:result:PResult} for
the observables $P_{4}^{\prime}$ and $P_5^{\prime}$. 
The observables $P_{6}^{\prime}$ and $P_{8}^{\prime}$ (as well as $S_7$ and $S_8$) are suppressed by the 
small size of the strong phase difference between the decay amplitudes, and therefore are expected to be 
close to zero across the whole $q^2$ region. 

In general, the measurements agree with SM expectations~\cite{Descotes-Genon:2013vna},
apart from a sizeable discrepancy in the interval \mbox{$4.30<q^2<8.68$\gevgevcccc} for the
observable $P_5^{\prime}$. The $p$-value, calculated using pseudo-experiments, with respect to the upper bound of the theoretical
  predictions given in Ref.~\cite{Descotes-Genon:2013vna}, for the observed deviation is $0.02\%$, 
corresponding to $3.7$ Gaussian standard deviations ($\sigma$). 
If we consider the 24 measurements as independent, 
the probability that at least one varies from the expected value by $3.7\, \sigma$ or more is approximately $0.5\%$. 
A discrepancy of $ 2.5 \, \sigma$ is observed integrating over the region \mbox{$1.0<q^2<6.0$\gevgevcccc} (see Table~\ref{tab:results}), 
which is considered the most robust region for theoretical predictions at large recoil. 
The discrepancy is also observed in the observable $S_5$. The value of $S_5$ quantifies the asymmetry between decays with positive and negative 
value of $\cos{\theta_K}$ for $|\phi|<\pi/2$, averaged with the opposite asymmetry of events with $|\phi|>\pi/2$~\cite{Altmannshofer:2008dz}. 
As a cross check, this asymmetry was also determined from a counting analysis. The result is consistent with the value for $S_{5}$ 
determined from the fit. 
It is worth noting that the predictions for the first two $q^2$-bins and
for the region \mbox{$1.0<q^2<6.0$\gevgevcccc} 
are also calculated in Ref.~\cite{Jager:2012uw}, where power corrections to the QCD factorization framework 
and resonance contributions are considered. 
However, there is not yet in the literature unanimous consensus about the best approach to treat these power corrections. 
The technique used in Ref.~\cite{Jager:2012uw} leads to a larger theoretical uncertainty with respect to Ref.~\cite{Descotes-Genon:2013vna}.

In conclusion, we measure for the first time the angular observables $S_4$, $S_5$, $S_7$, $S_8$ and the 
corresponding form-factor independent observables $P_{4}^{\prime}$, $P_5^{\prime}$, $P_6^{\prime}$ and 
$P_8^{\prime}$ in the decay \BdToKstmm. 
These measurements have been performed in six $q^2$ bins for each of the four observables. 
Agreement with SM predictions~\cite{Descotes-Genon:2013vna} is observed for 23 of the 24 measurements, 
while a local discrepancy of $3.7\, \sigma$ is observed in the interval \mbox{$4.30<q^2<8.68$\gevgevcccc} for the observable $P_5^{\prime}$.   
Integrating over the region \mbox{$1.0<q^2<6.0$\gevgevcccc}, the observed discrepancy in $P_5^{\prime}$ is $ 2.5 \, \sigma$.
The observed discrepancy in the angular observable $P_{5}^{\prime}$ could be caused by a smaller value of the Wilson coefficient $C_9$ 
with respect to the SM, as has been suggested to explain some other small inconsistencies between the $\Bd \to \Kstarz\mumu$ 
data~\cite{Aaij:2013iag} and SM predictions~\cite{Descotes-Genon:2013wba}.
Measurements with more data  
and further theoretical studies will be important to draw more definitive conclusions about this discrepancy.

\subsubsection*{Acknowledgements}

\noindent We express our gratitude to our colleagues in the CERN
accelerator departments for the excellent performance of the LHC. We
thank the technical and administrative staff at the LHCb
institutes. We acknowledge support from CERN and from the national
agencies: CAPES, CNPq, FAPERJ and FINEP (Brazil); NSFC (China);
CNRS/IN2P3 and Region Auvergne (France); BMBF, DFG, HGF and MPG
(Germany); SFI (Ireland); INFN (Italy); FOM and NWO (The Netherlands);
SCSR (Poland); MEN/IFA (Romania); MinES, Rosatom, RFBR and NRC
``Kurchatov Institute'' (Russia); MinECo, XuntaGal and GENCAT (Spain);
SNSF and SER (Switzerland); NAS Ukraine (Ukraine); STFC (United
Kingdom); NSF (USA). We also acknowledge the support received from the
ERC under FP7. The Tier1 computing centres are supported by IN2P3
(France), KIT and BMBF (Germany), INFN (Italy), NWO and SURF (The
Netherlands), PIC (Spain), GridPP (United Kingdom). We are thankful
for the computing resources put at our disposal by Yandex LLC
(Russia), as well as to the communities behind the multiple open
source software packages that we depend on.

\addcontentsline{toc}{section}{References}
\setboolean{inbibliography}{true}
\bibliographystyle{LHCb}
\bibliography{main,rare,stat,LHCb-PAPER,LHCb-CONF,LHCb-DP}

\ifx\mcitethebibliography\mciteundefinedmacro
\PackageError{LHCb.bst}{mciteplus.sty has not been loaded}
{This bibstyle requires the use of the mciteplus package.}\fi
\providecommand{\href}[2]{#2}
\begin{mcitethebibliography}{10}
\mciteSetBstSublistMode{n}
\mciteSetBstMaxWidthForm{subitem}{\alph{mcitesubitemcount})}
\mciteSetBstSublistLabelBeginEnd{\mcitemaxwidthsubitemform\space}
{\relax}{\relax}

\bibitem{Altmannshofer:2008dz}
W.~Altmannshofer {\em et~al.}, \ifthenelse{\boolean{articletitles}}{{\it
  {Symmetries and asymmetries of $B \to K^{*} \mu^{+} \mu^{-}$ decays in the
  Standard Model and beyond}},
  }{}\href{http://dx.doi.org/10.1088/1126-6708/2009/01/019}{JHEP {\bf 01}
  (2009) 019}, \href{http://arxiv.org/abs/0811.1214}{{\tt
  arXiv:0811.1214}}\relax
\mciteBstWouldAddEndPuncttrue
\mciteSetBstMidEndSepPunct{\mcitedefaultmidpunct}
{\mcitedefaultendpunct}{\mcitedefaultseppunct}\relax
\EndOfBibitem
\bibitem{Becirevic:2011bp}
D.~Be\v{c}irevi\'{c} and E.~Schneider,
  \ifthenelse{\boolean{articletitles}}{{\it {On transverse asymmetries in $B\to
  K^{*}\ell \ell$}},
  }{}\href{http://dx.doi.org/10.1016/j.nuclphysb.2011.09.004}{Nucl.\ Phys.\
  {\bf B854} (2012) 321}, \href{http://arxiv.org/abs/1106.3283}{{\tt
  arXiv:1106.3283}}\relax
\mciteBstWouldAddEndPuncttrue
\mciteSetBstMidEndSepPunct{\mcitedefaultmidpunct}
{\mcitedefaultendpunct}{\mcitedefaultseppunct}\relax
\EndOfBibitem
\bibitem{Matias:2012xw}
J.~Matias, F.~Mescia, M.~Ramon, and J.~Virto,
  \ifthenelse{\boolean{articletitles}}{{\it {Complete anatomy of $\Bzb \to
  \Kstarzb (\to K \pi) \ell^+\ell^-$ and its angular distribution}},
  }{}\href{http://dx.doi.org/10.1007/JHEP04(2012)104}{JHEP {\bf 04} (2012)
  104}, \href{http://arxiv.org/abs/1202.4266}{{\tt arXiv:1202.4266}}\relax
\mciteBstWouldAddEndPuncttrue
\mciteSetBstMidEndSepPunct{\mcitedefaultmidpunct}
{\mcitedefaultendpunct}{\mcitedefaultseppunct}\relax
\EndOfBibitem
\bibitem{Aubert:2008ju}
{BaBar collaboration}, B.~Aubert {\em et~al.},
  \ifthenelse{\boolean{articletitles}}{{\it {Angular distributions in the decay
  $B \to K^*\ell^+\ell^-$}},
  }{}\href{http://dx.doi.org/10.1103/PhysRevD.79.031102}{Phys.\ \, Rev.\  {\bf
  D79} (2009) 031102}, \href{http://arxiv.org/abs/0804.4412}{{\tt
  arXiv:0804.4412}}\relax
\mciteBstWouldAddEndPuncttrue
\mciteSetBstMidEndSepPunct{\mcitedefaultmidpunct}
{\mcitedefaultendpunct}{\mcitedefaultseppunct}\relax
\EndOfBibitem
\bibitem{Wei:2009zv}
Belle collaboration, J.-T. Wei {\em et~al.},
  \ifthenelse{\boolean{articletitles}}{{\it {Measurement of the differential
  branching fraction and forward-backward asymmetry for $B \to K^{(*)}
  l^+l^-$}}, }{}\href{http://dx.doi.org/10.1103/PhysRevLett.103.171801}{Phys.\
  Rev.\ Lett.\  {\bf 103} (2009) 171801},
  \href{http://arxiv.org/abs/0904.0770}{{\tt arXiv:0904.0770}}\relax
\mciteBstWouldAddEndPuncttrue
\mciteSetBstMidEndSepPunct{\mcitedefaultmidpunct}
{\mcitedefaultendpunct}{\mcitedefaultseppunct}\relax
\EndOfBibitem
\bibitem{Aaltonen:2011ja}
CDF collaboration, T.~Aaltonen {\em et~al.},
  \ifthenelse{\boolean{articletitles}}{{\it {Measurements of the angular
  distributions in the decays $B \to K^{(*)} \mu^+ \mu^-$ at CDF}},
  }{}\href{http://dx.doi.org/10.1103/PhysRevLett.108.081807}{Phys.\ Rev.\
  Lett.\  {\bf 108} (2012) 081807}, \href{http://arxiv.org/abs/1108.0695}{{\tt
  arXiv:1108.0695}}\relax
\mciteBstWouldAddEndPuncttrue
\mciteSetBstMidEndSepPunct{\mcitedefaultmidpunct}
{\mcitedefaultendpunct}{\mcitedefaultseppunct}\relax
\EndOfBibitem
\bibitem{Aaij:2013iag}
LHCb collaboration, R.~Aaij {\em et~al.},
  \ifthenelse{\boolean{articletitles}}{{\it {Differential branching fraction
  and angular analysis of the decay $B^{0} \rightarrow K^{*0}
  \mu^{+}\mu^{-}$}}, }{}\href{http://arxiv.org/abs/1304.6325}{{\tt
  arXiv:1304.6325}}, {submitted to JHEP}\relax
\mciteBstWouldAddEndPuncttrue
\mciteSetBstMidEndSepPunct{\mcitedefaultmidpunct}
{\mcitedefaultendpunct}{\mcitedefaultseppunct}\relax
\EndOfBibitem
\bibitem{Kruger:2005ep}
F.~Kruger and J.~Matias, \ifthenelse{\boolean{articletitles}}{{\it {Probing new
  physics via the transverse amplitudes of $B^0 \to K^{*0} (\to K^- \pi^+)
  \ell^+\ell^-$ at large recoil}},
  }{}\href{http://dx.doi.org/10.1103/PhysRevD.71.094009}{Phys.\ Rev.\  {\bf
  D71} (2005) 094009}, \href{http://arxiv.org/abs/hep-ph/0502060}{{\tt
  arXiv:hep-ph/0502060}}\relax
\mciteBstWouldAddEndPuncttrue
\mciteSetBstMidEndSepPunct{\mcitedefaultmidpunct}
{\mcitedefaultendpunct}{\mcitedefaultseppunct}\relax
\EndOfBibitem
\bibitem{Egede:2008uy}
U.~Egede {\em et~al.}, \ifthenelse{\boolean{articletitles}}{{\it {New
  observables in the decay mode $\Bzb \to \Kstarzb \ell^+ \ell^-$}},
  }{}\href{http://dx.doi.org/10.1088/1126-6708/2008/11/032}{JHEP {\bf 11}
  (2008) 032}, \href{http://arxiv.org/abs/0807.2589}{{\tt
  arXiv:0807.2589}}\relax
\mciteBstWouldAddEndPuncttrue
\mciteSetBstMidEndSepPunct{\mcitedefaultmidpunct}
{\mcitedefaultendpunct}{\mcitedefaultseppunct}\relax
\EndOfBibitem
\bibitem{Bobeth:2011gi}
C.~Bobeth, G.~Hiller, and D.~van Dyk, \ifthenelse{\boolean{articletitles}}{{\it
  {More benefits of semileptonic rare $B$ decays at low recoil: CP violation}},
  }{}\href{http://dx.doi.org/10.1007/JHEP07(2011)067}{JHEP {\bf 07} (2011)
  067}, \href{http://arxiv.org/abs/1105.0376}{{\tt arXiv:1105.0376}}\relax
\mciteBstWouldAddEndPuncttrue
\mciteSetBstMidEndSepPunct{\mcitedefaultmidpunct}
{\mcitedefaultendpunct}{\mcitedefaultseppunct}\relax
\EndOfBibitem
\bibitem{Descotes-Genon:2013vna}
S.~Descotes-Genon, T.~Hurth, J.~Matias, and J.~Virto,
  \ifthenelse{\boolean{articletitles}}{{\it {Optimizing the basis of ${B} \to
  {K}^{*}\ell^+ \ell^-$ observables in the full kinematic range}},
  }{}\href{http://dx.doi.org/10.1007/JHEP05(2013)137}{JHEP {\bf 05} (2013)
  137}, \href{http://arxiv.org/abs/1303.5794}{{\tt arXiv:1303.5794}}\relax
\mciteBstWouldAddEndPuncttrue
\mciteSetBstMidEndSepPunct{\mcitedefaultmidpunct}
{\mcitedefaultendpunct}{\mcitedefaultseppunct}\relax
\EndOfBibitem
\bibitem{Alves:2008zz}
LHCb collaboration, A.~A. Alves~Jr. {\em et~al.},
  \ifthenelse{\boolean{articletitles}}{{\it {The \lhcb detector at the LHC}},
  }{}\href{http://dx.doi.org/10.1088/1748-0221/3/08/S08005}{JINST {\bf 3}
  (2008) S08005}\relax
\mciteBstWouldAddEndPuncttrue
\mciteSetBstMidEndSepPunct{\mcitedefaultmidpunct}
{\mcitedefaultendpunct}{\mcitedefaultseppunct}\relax
\EndOfBibitem
\bibitem{Adinolfi:1495721}
M.~Adinolfi {\em et~al.}, \ifthenelse{\boolean{articletitles}}{{\it
  {Performance of the \lhcb RICH detector at the LHC}}, }{}Eur.\ Phys.\ J.\
  {\bf C73} (2012) 2431\relax
\mciteBstWouldAddEndPuncttrue
\mciteSetBstMidEndSepPunct{\mcitedefaultmidpunct}
{\mcitedefaultendpunct}{\mcitedefaultseppunct}\relax
\EndOfBibitem
\bibitem{AlvesJr:1492807}
A.~A. Alves~Jr {\em et~al.}, \ifthenelse{\boolean{articletitles}}{{\it
  {Performance of the \lhcb muon system}}, }{}JINST {\bf 8} (2012) P02022\relax
\mciteBstWouldAddEndPuncttrue
\mciteSetBstMidEndSepPunct{\mcitedefaultmidpunct}
{\mcitedefaultendpunct}{\mcitedefaultseppunct}\relax
\EndOfBibitem
\bibitem{Aaij:2012me}
R.~Aaij {\em et~al.}, \ifthenelse{\boolean{articletitles}}{{\it {The \lhcb
  trigger and its performance in 2011}},
  }{}\href{http://dx.doi.org/10.1088/1748-0221/8/04/P04022}{JINST {\bf 8}
  (2013) P04022}, \href{http://arxiv.org/abs/1211.3055}{{\tt
  arXiv:1211.3055}}\relax
\mciteBstWouldAddEndPuncttrue
\mciteSetBstMidEndSepPunct{\mcitedefaultmidpunct}
{\mcitedefaultendpunct}{\mcitedefaultseppunct}\relax
\EndOfBibitem
\bibitem{Sjostrand:2006za}
T.~Sj\"{o}strand, S.~Mrenna, and P.~Skands,
  \ifthenelse{\boolean{articletitles}}{{\it {PYTHIA 6.4 physics and manual}},
  }{}\href{http://dx.doi.org/10.1088/1126-6708/2006/05/026}{JHEP {\bf 05}
  (2006) 026}, \href{http://arxiv.org/abs/hep-ph/0603175}{{\tt
  arXiv:hep-ph/0603175}}\relax
\mciteBstWouldAddEndPuncttrue
\mciteSetBstMidEndSepPunct{\mcitedefaultmidpunct}
{\mcitedefaultendpunct}{\mcitedefaultseppunct}\relax
\EndOfBibitem
\bibitem{LHCb-PROC-2010-056}
I.~Belyaev {\em et~al.}, \ifthenelse{\boolean{articletitles}}{{\it {Handling of
  the generation of primary events in \gauss, the \lhcb simulation framework}},
  }{}\href{http://dx.doi.org/10.1109/NSSMIC.2010.5873949}{Nuclear Science
  Symposium Conference Record (NSS/MIC) {\bf IEEE} (2010) 1155}\relax
\mciteBstWouldAddEndPuncttrue
\mciteSetBstMidEndSepPunct{\mcitedefaultmidpunct}
{\mcitedefaultendpunct}{\mcitedefaultseppunct}\relax
\EndOfBibitem
\bibitem{Lange:2001uf}
D.~Lange, \ifthenelse{\boolean{articletitles}}{{\it {The EvtGen particle decay
  simulation package}},
  }{}\href{http://dx.doi.org/10.1016/S0168-9002(01)00089-4}{Nucl.\ Instrum.\
  Meth.\  {\bf A462} (2001) 152}\relax
\mciteBstWouldAddEndPuncttrue
\mciteSetBstMidEndSepPunct{\mcitedefaultmidpunct}
{\mcitedefaultendpunct}{\mcitedefaultseppunct}\relax
\EndOfBibitem
\bibitem{Golonka:2005pn}
P.~Golonka and Z.~Was, \ifthenelse{\boolean{articletitles}}{{\it {PHOTOS Monte
  Carlo: A Precision tool for QED corrections in $Z$ and $W$ decays}},
  }{}\href{http://dx.doi.org/10.1140/epjc/s2005-02396-4}{Eur.\ Phys.\ J.\  {\bf
  C45} (2006) 97}, \href{http://arxiv.org/abs/hep-ph/0506026}{{\tt
  arXiv:hep-ph/0506026}}\relax
\mciteBstWouldAddEndPuncttrue
\mciteSetBstMidEndSepPunct{\mcitedefaultmidpunct}
{\mcitedefaultendpunct}{\mcitedefaultseppunct}\relax
\EndOfBibitem
\bibitem{Allison:2006ve}
J.~Allison {\em et~al.}, \ifthenelse{\boolean{articletitles}}{{\it {Geant4
  developments and applications}},
  }{}\href{http://dx.doi.org/10.1109/TNS.2006.869826}{IEEE Trans.\ Nucl.\ Sci.\
   {\bf 53} (2006) 270}\relax
\mciteBstWouldAddEndPuncttrue
\mciteSetBstMidEndSepPunct{\mcitedefaultmidpunct}
{\mcitedefaultendpunct}{\mcitedefaultseppunct}\relax
\EndOfBibitem
\bibitem{Agostinelli:2002hh}
GEANT collaboration, S.~Agostinelli {\em et~al.},
  \ifthenelse{\boolean{articletitles}}{{\it {GEANT4 - A simulation toolkit}},
  }{}\href{http://dx.doi.org/10.1016/S0168-9002(03)01368-8}{Nucl.\ Instrum.\
  Meth.\  {\bf A506} (2003) 250}\relax
\mciteBstWouldAddEndPuncttrue
\mciteSetBstMidEndSepPunct{\mcitedefaultmidpunct}
{\mcitedefaultendpunct}{\mcitedefaultseppunct}\relax
\EndOfBibitem
\bibitem{LHCb-PROC-2011-006}
M.~Clemencic {\em et~al.}, \ifthenelse{\boolean{articletitles}}{{\it {The \lhcb
  simulation application, \gauss: design, evolution and experience}},
  }{}\href{http://dx.doi.org/10.1088/1742-6596/331/3/032023}{{J.\ of Phys:
  Conf.\ Ser.\ } {\bf 331} (2011) 032023}\relax
\mciteBstWouldAddEndPuncttrue
\mciteSetBstMidEndSepPunct{\mcitedefaultmidpunct}
{\mcitedefaultendpunct}{\mcitedefaultseppunct}\relax
\EndOfBibitem
\bibitem{Breiman}
L.~Breiman, J.~H. Friedman, R.~A. Olshen, and C.~J. Stone, {\em Classification
  and regression trees}, Wadsworth international group, Belmont, California,
  USA, 1984\relax
\mciteBstWouldAddEndPuncttrue
\mciteSetBstMidEndSepPunct{\mcitedefaultmidpunct}
{\mcitedefaultendpunct}{\mcitedefaultseppunct}\relax
\EndOfBibitem
\bibitem{AdaBoost}
Y.~Freund and R.~E. Schapire, \ifthenelse{\boolean{articletitles}}{{\it A
  decision-theoretic generalization of on-line learning and an application to
  boosting}, }{}\href{http://dx.doi.org/10.1006/jcss.1997.1504}{Jour.\ Comp.\
  and Syst.\ Sc.\  {\bf 55} (1997) 119}\relax
\mciteBstWouldAddEndPuncttrue
\mciteSetBstMidEndSepPunct{\mcitedefaultmidpunct}
{\mcitedefaultendpunct}{\mcitedefaultseppunct}\relax
\EndOfBibitem
\bibitem{michelThesis}
M.~De~Cian, {\em Track reconstruction efficiency and analysis of
  $B^{0}\rightarrow K^{*0}\mu^{+}\mu^{-}$ at the LHCb experiment}, PhD thesis,
  University of Zurich, 2013\relax
\mciteBstWouldAddEndPuncttrue
\mciteSetBstMidEndSepPunct{\mcitedefaultmidpunct}
{\mcitedefaultendpunct}{\mcitedefaultseppunct}\relax
\EndOfBibitem
\bibitem{Skwarnicki:1986xj}
T.~Skwarnicki, {\em {A study of the radiative cascade transitions between the
  Upsilon-prime and Upsilon resonances}}, PhD thesis, Institute of Nuclear
  Physics, Krakow, 1986,
  {\href{http://inspirehep.net/record/230779}{DESY-F31-86-02}}\relax
\mciteBstWouldAddEndPuncttrue
\mciteSetBstMidEndSepPunct{\mcitedefaultmidpunct}
{\mcitedefaultendpunct}{\mcitedefaultseppunct}\relax
\EndOfBibitem
\bibitem{Matias:2012qz}
J.~Matias, \ifthenelse{\boolean{articletitles}}{{\it {On the S-wave pollution
  of $B \to K^{*} l^{+} l^{-}$ observables}},
  }{}\href{http://dx.doi.org/10.1103/PhysRevD.86.094024}{Phys.\ Rev.\  {\bf
  D86} (2012) 094024}, \href{http://arxiv.org/abs/1209.1525}{{\tt
  arXiv:1209.1525}}\relax
\mciteBstWouldAddEndPuncttrue
\mciteSetBstMidEndSepPunct{\mcitedefaultmidpunct}
{\mcitedefaultendpunct}{\mcitedefaultseppunct}\relax
\EndOfBibitem
\bibitem{Feldman:1997qc}
G.~J. Feldman and R.~D. Cousins, \ifthenelse{\boolean{articletitles}}{{\it
  {Unified approach to the classical statistical analysis of small signals}},
  }{}\href{http://dx.doi.org/10.1103/PhysRevD.57.3873}{Phys.\ Rev.\  {\bf D57}
  (1998) 3873}, \href{http://arxiv.org/abs/physics/9711021}{{\tt
  arXiv:physics/9711021}}\relax
\mciteBstWouldAddEndPuncttrue
\mciteSetBstMidEndSepPunct{\mcitedefaultmidpunct}
{\mcitedefaultendpunct}{\mcitedefaultseppunct}\relax
\EndOfBibitem
\bibitem{Jager:2012uw}
S.~J{\"{a}}ger and J.~M. Camalich, \ifthenelse{\boolean{articletitles}}{{\it
  {On $B \rightarrow Vll$ at small dilepton invariant mass, power corrections,
  and new physics}}, }{}\href{http://dx.doi.org/10.1007/JHEP05(2013)043}{JHEP
  {\bf 05} (2013) 043}, \href{http://arxiv.org/abs/1212.2263}{{\tt
  arXiv:1212.2263}}\relax
\mciteBstWouldAddEndPuncttrue
\mciteSetBstMidEndSepPunct{\mcitedefaultmidpunct}
{\mcitedefaultendpunct}{\mcitedefaultseppunct}\relax
\EndOfBibitem
\bibitem{Descotes-Genon:2013wba}
S.~Descotes-Genon, J.~Matias, and J.~Virto,
  \ifthenelse{\boolean{articletitles}}{{\it {Understanding the $\Bd \to
  \Kstarz\mumu$ Anomaly}}, }{}\href{http://arxiv.org/abs/1307.5683}{{\tt
  arXiv:1307.5683}}\relax
\mciteBstWouldAddEndPuncttrue
\mciteSetBstMidEndSepPunct{\mcitedefaultmidpunct}
{\mcitedefaultendpunct}{\mcitedefaultseppunct}\relax
\EndOfBibitem
\end{mcitethebibliography}

\clearpage

\end{document}